\documentclass[10pt,letterpaper,compsoc,conference]{iiswc26}

\usepackage{cite}
\usepackage{amsmath,amssymb,amsfonts}
\usepackage{algorithmic}
\usepackage{graphicx}
\usepackage[dvipsnames]{xcolor}
\usepackage[final]{microtype}
\usepackage[italic]{mathastext}
\usepackage{libertine}
\usepackage[T1]{fontenc}
\usepackage{textcomp}
\usepackage[varqu,varl]{zi4}
\usepackage[all]{nowidow}
\usepackage[keeplastbox]{flushend}
\usepackage{fancyhdr}

\usepackage{booktabs}
\usepackage{multirow}
\usepackage{tabularx}
\usepackage{caption}
\usepackage{colortbl}  
\usepackage{siunitx} 
\usepackage{threeparttable}
\usepackage{changes}
\usepackage{authblk}
\usetikzlibrary{fit}

\fancypagestyle{firstpage}{
  \fancyhf{}
  
}

\begin{document}


\title{CARB: A Characterization-Guided Framework for CNN Inference Cost Prediction and Deployment Screening}





\author{Linh Nguyen, Zhixin Pan}
\affil{Florida State University}

\maketitle
\thispagestyle{firstpage}
\pagestyle{plain}


\begin{abstract}
Accurate pre-deployment estimation of CNN inference cost---energy, latency, and peak memory---is increasingly critical as models are deployed on resource-constrained GPU platforms. Existing approaches rely on FLOPs, latency measurements, or single-device profiling as energy proxies, overlooking the non-linear interactions between architectural design and hardware load. We present a workload characterization study of \num{13419} CNN configurations on two GPU platforms (RTX\,5090 and RTX\,3080) under GPU telemetry, revealing that energy, latency, and memory exhibit fundamentally distinct scaling behaviors: energy and latency diverge by $3\times$ under high computational demand, and cross-GPU transferability differs by target---energy and latency require platform-specific models while memory transfers well across the two tested platforms. Building on these characterization findings, we develop CARB, a cascade-blended ensemble that jointly predicts all three targets with $R^2\,{\approx}\,0.99$, and a two-stage deployment screening workflow that eliminates over 90\% of candidates in seconds, reducing large design spaces to a Pareto-prioritized shortlist validated against real hardware.
\end{abstract}

\section{Introduction}

The growing deployment of convolutional neural networks (CNNs)~\cite{krizhevsky2012imagenet} on GPU platforms has increased the need for accurate pre-deployment estimation of inference costs~\cite{canziani2016analysis}. Energy consumption, inference latency, and peak memory each impose distinct constraints~\cite{strubell2019energy} on model selection and hardware provisioning, yet they exhibit fundamentally different scaling behaviors that existing estimation approaches fail to capture jointly.

Current practice typically relies on FLOPs, latency measurements, or single-device profiling as proxies for energy. FLOPs-based estimates treat all operations as equally expensive, ignoring how architectural choices and hardware execution efficiency interact. Latency-based proxies assume energy and latency scale proportionally, an assumption that breaks down under high computational demand. Single-device profiles assume cross-GPU transferability, overlooking that energy and latency scale differently across hardware generations; memory, by contrast, transfers well across the two platforms we tested. Each of these shortcuts introduces systematic errors that propagate into deployment decisions.

In this paper, we take a characterization-guided approach. We begin with a large-scale empirical study of \num{13419} CNN configurations on two GPU platforms (RTX\,5090 and RTX\,3080), which reveals several findings that fundamentally challenge the use of existing proxies. Most critically, energy and latency diverge substantially under high computational demand: energy scales $35.4\times$ across batch-size regimes while latency scales only $11.2\times$, making latency an unreliable energy proxy. Cross-GPU analysis further reveals that energy and latency follow non-unit transfer slopes ($1.61\times$ and $2.09\times$) while memory transfers with slope ${\approx}1$ across the two tested platforms---a target-level asymmetry that directly shapes the design of our prediction framework.


Motivated by these findings, we propose a two-stage design-space exploration framework that uses target-aware per-GPU ensemble models to jointly predict energy, latency, and memory, reducing large configuration search spaces to Pareto-prioritized candidates before any hardware profiling. These contributions demonstrate the necessity of direct energy modeling and enable practical energy-aware CNN deployment with significantly reduced profiling cost.

Our contributions are organized around three core characterization findings and their practical implications:

\begin{itemize}
  \item \textbf{Energy, latency, and memory are not interchangeable.} Under high computational demand, energy scales $35.4\times$ while latency scales only $11.2\times$ across batch-size regimes. FP16 reduces memory by $1.7\times$ but energy by only $1.1\times$. No single architectural knob determines cost in isolation.
 
  \item \textbf{Cross-GPU transferability is target-dependent.} Energy and latency follow non-unit transfer slopes ($1.61\times$, $2.09\times$) and require platform-specific models, while memory transfers with slope ${\approx}1$ across the two tested platforms.
 
  \item \textbf{These insights enable efficient deployment screening.} CARB, a cascade ensemble guided by the above findings, supports a two-stage screening workflow that reduces a \num{3072}-configuration space to a Pareto-prioritized shortlist before any hardware profiling, with accuracy ($R^2\,{\approx}\,0.99$) sufficient for reliable budget classification.
\end{itemize}

\section{Related Work}

A substantial body of work has addressed the prediction of CNN inference costs on hardware platforms. NeuralPower~\cite{cai2017neuralpower} uses sparse polynomial regression to predict per-layer power, runtime, and energy consumption across GPU platforms. Hardware-aware neural architecture search (NAS) algorithms~\cite{cai2018proxylessnas, wu2019fbnet} embed latency lookup tables into architecture search, while nn-Meter~\cite{zhang2021nnmeter} predicts latency on diverse edge devices. MAPLE~\cite{abbasi2022maple} generalizes latency prediction to unseen devices with few-shot adaptation, and NeuSight~\cite{lee2025forecasting} forecasts GPU performance across multiple platforms using operator-level predictors. While these works establish the viability of data-driven cost prediction, the NAS-oriented and device-generalization methods focus predominantly on latency, and none characterizes the divergence between energy and latency behavior at scale.

The relationship between energy and latency has received some attention. Prior work on multi-GPU energy profiling finds that GPU generation can produce up to 40\% energy differences for the same CNN workload~\cite{castro2019energy}. DVFS-aware models~\cite{han2025dvfs} show that frequency scaling affects latency and energy differently, motivating energy as an independent metric. These works characterize energy differences across GPU generations or frequency settings, but do not quantify how energy and latency diverge under computational demand; we show this gap reaches $35.4\times$ vs.\ $11.2\times$ across batch sizes under high GPU utilization.

Cross-device performance prediction has been explored primarily for latency~\cite{abbasi2022maple, lee2025forecasting}. Cross-GPU energy transfer has received little attention. We show that energy and latency follow non-unit cross-GPU slopes ($1.61\times$ and $2.09\times$), ruling out fixed-multiplier transfer, while memory transfers with slope ${\approx}1$ across our two tested platforms---a target-level asymmetry that directly informs our model design.

Most existing prediction work targets a single metric; joint prediction of energy, latency, and memory is rare. HyperPower~\cite{stamoulis2018hyperpower} jointly optimizes power and memory during architecture search but does not provide a deployment prediction model. Latenrgy~\cite{pittman2024latenrgy} predicts latency and energy but omits memory. Our work jointly predicts all three targets, treats each target's cross-GPU transferability differently, and addresses practical pre-deployment screening---identifying feasible configurations against a hardware budget before any profiling is run---a gap that prior work does not fill. Prior IISWC studies characterize deep learning workloads and accelerator behavior~\cite{rodrigues2017fine, siu2018memory, wang2019characterizing}, but do not provide systematic inference-cost characterization across a broad architectural design space.
\section{Data Collection}

We construct a dataset by benchmarking CNN configurations under controlled hardware conditions on two GPU platforms: NVIDIA RTX\,5090 and RTX\,3080. The collection procedure consists of three steps: GPU environment control, configuration sampling, and unified measurement.

\subsection{Hardware and Software Setup}

To reduce measurement variance caused by dynamic voltage and frequency scaling (DVFS), we lock GPU clocks before each collection run (graphics clock: 1350\,MHz; memory clock: 5001\,MHz). cuDNN benchmark mode is disabled to avoid run-to-run kernel selection variance. Between configurations, we apply Python garbage collection, CUDA cache clearing, and device synchronization, followed by 1-second cooling interval to allow the GPU to stabilize. All experiments use the same PyTorch~\cite{paszke2019pytorch} version (v.2.8.0) across both platforms. 

\subsection{Search Space}

We use a ResNet-style search space~\cite{he2016deep} that covers both basic-block and bottleneck-block variants of the architecture. The two block types differ structurally---basic blocks consist of two $3{\times}3$ convolutions, while bottleneck blocks use a $1{\times}1$--$3{\times}3$--$1{\times}1$ design with expansion factor\,4---and therefore admit different depth configurations (Table~\ref{tab:search_space}). Configurations are randomly shuffled before measurement to avoid systematic ordering effects on the reported metrics, while configurations exceeding GPU memory capacity are skipped and recorded as out-of-memory (OOM) entries.

\begin{table}[t]
\centering
\caption{Search space used for dataset generation. Basic and bottleneck block types use disjoint depth sets due to their structural differences.}
\label{tab:search_space}
\small
\begin{tabular}{l l}
\hline
Factor & Values \\
\hline
Block type & \{Basic, Bottleneck\} \\
Depth (Basic) & \{8, 14, 20, 32, 44, 56, 110, 152\} \\
Depth (Bottleneck) & \{50, 101, 152, 200\} \\
Width multiplier & \{0.1, 0.25, 0.5, 1.0, 1.5, 2.0, 3.0, 4.0\} \\
Early downsampling & \{True, False\} \\
Input resolution & \{32, 64, 96, 128\} \\
Batch size & \{1, 2, 4, 8, 16, 32, 64, 128, 256\} \\
Precision & \{FP32, FP16\} \\
\hline
\end{tabular}
\end{table}

\subsection{Measurement}

For each configuration, we instantiate the model, generate a synthetic input tensor of shape $(B, 3, H, W)$, and run 10 warmup passes to stabilize kernel caching before recording measurements. No other GPU workloads run concurrently during collection. Latency is the mean inference time per batch over 100 runs in evaluation mode after 10 warmup iterations. Energy is recorded as the difference in the NVIDIA Management Library (NVML) cumulative energy counter before and after the inference loop, converted from millijoules to joules. Peak memory is measured via \text{max\_memory\_allocated} after calling \text{reset\_peak\_memory\_stats} before each forward pass to ensure isolation from prior operations. GPU temperature is recorded as the average of readings taken immediately before and after the inference loop. FLOPs are computed using \text{fvcore} \text{FlopCountAnalysis}~\cite{fvcore2020}. Measurement variance is reduced through three complementary mechanisms: GPU clock locking (eliminating DVFS-induced fluctuations), the inter-configuration cooling interval (allowing thermal stabilization), and averaging over 100 inference runs (reducing transient noise).

GPU utilization (SM and memory bandwidth utilization) is sampled using NVML over a 2-second window during a training-mode forward-backward pass, and kernel launch counts are obtained from the PyTorch profiler on a single forward pass. These runtime features support workload characterization and interpretability; the primary predictive signal comes from architectural and derived static features.

\subsection{Collected Features}

\begin{table}[t]
\centering
\caption{Collected features for prediction and analysis.}
\label{tab:collected_features}
\small
\begin{tabularx}{\columnwidth}{l X}
\hline
Category & Features \\
\hline
Architectural & Depth, block type, width multiplier, precision, batch size, input size, early downsampling \\
Derived static & \#Conv/BN/activation/FC layers, max channel width, \#downsamples, FLOPs, \#params, $M_{\text{param}}$, $M_{\text{activation}}$, $M_{\text{peak}}$, $AI$ \\
Runtime & SM utilization, memory utilization, GPU temperature, kernel launches \\
Targets & Energy\,(J), Latency\,(ms), Peak memory\,(MB) \\
\hline
\end{tabularx}
\end{table}

We collect a comprehensive set of features from three categories, summarized in Table~\ref{tab:collected_features}. Activation memory is estimated using forward hooks by summing the byte sizes of all layer outputs. We also define an arithmetic intensity proxy $AI = \text{FLOPs} / (M_{\text{param}} + M_{\text{activation}})$~\cite{williams2009roofline}, which captures the ratio of computation to memory traffic and reflects differences in hardware utilization across architectures.
\section{Workload and Energy Characterization}
\label{sec:analysis}

In this section, we conduct a systematic empirical analysis to characterize how architectural design choices affect energy, latency, and peak memory, and to establish where energy diverges from latency as a cost metric. Unless otherwise noted, all analyses in this section use the RTX\,5090 dataset; the cross-GPU comparison in Section~\ref{sec:crossgpu} additionally draws on the RTX\,3080 dataset.

\begin{figure*}[h]
\centering
    \includegraphics[width=\textwidth]{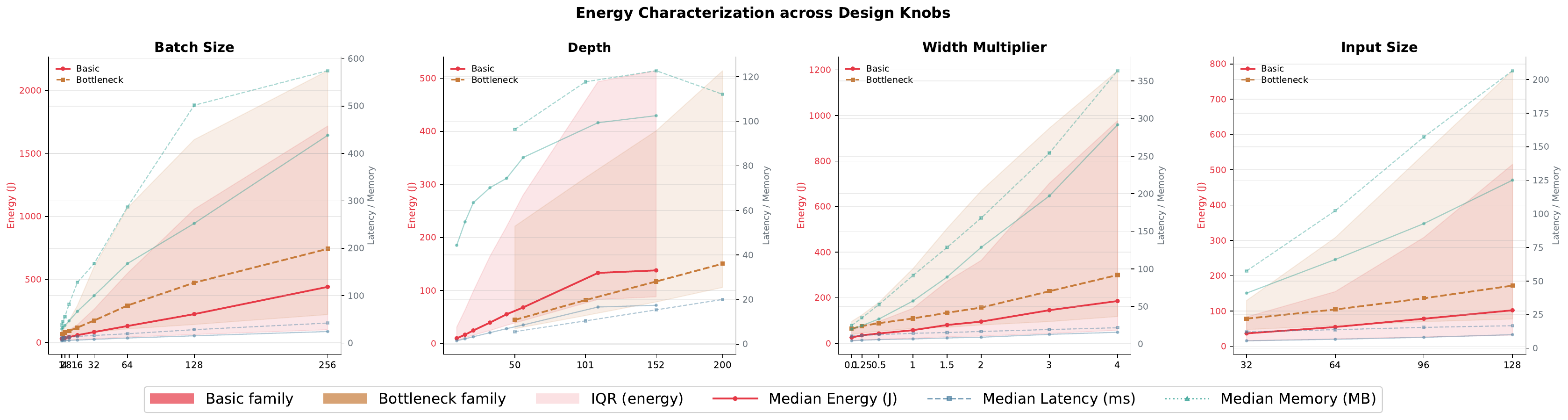}
    \caption{Energy characterization across design knobs. Each panel shows the basic family (solid red line) and the bottleneck family (dashed orange line) separately. Primary axis (left): Energy (J); secondary axis (right, faded): Latency (ms) and Peak Memory (MB).}
    \label{fig:char}
\end{figure*}


\begin{figure*}[h]
\centering
    \includegraphics[width=\textwidth]{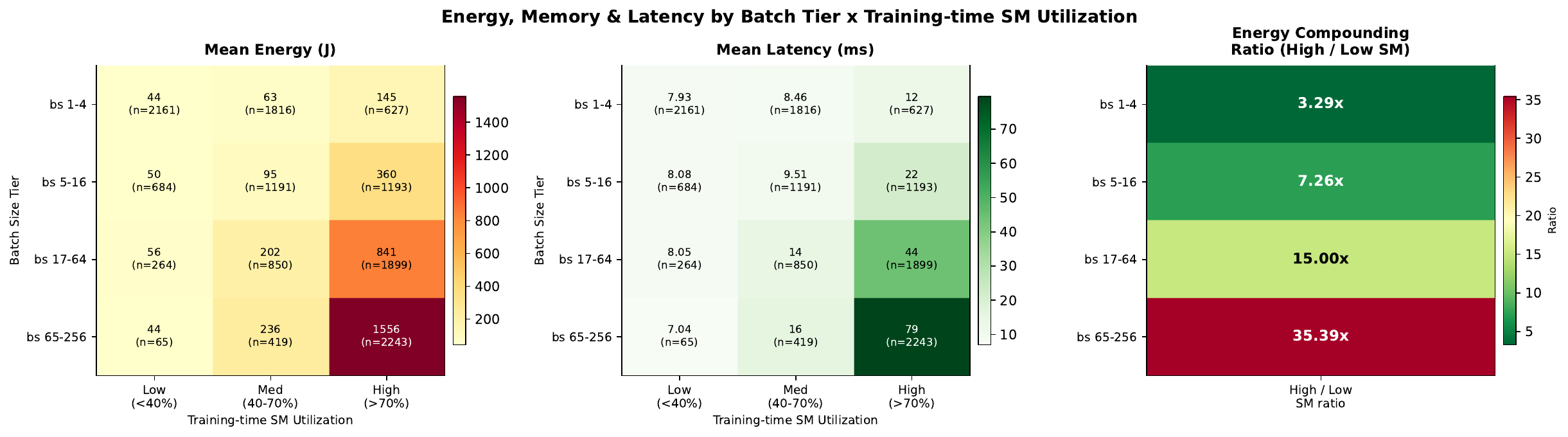}
    \caption{Mean energy and latency across batch-size tiers and GPU SM utilization levels (measured during training). The rightmost panel shows the relative energy scaling (High-SM / Low-SM) per batch tier.}
    \label{fig:crossfactor}
\end{figure*}

\begin{figure*}[h]
\centering
    \includegraphics[width=\textwidth]{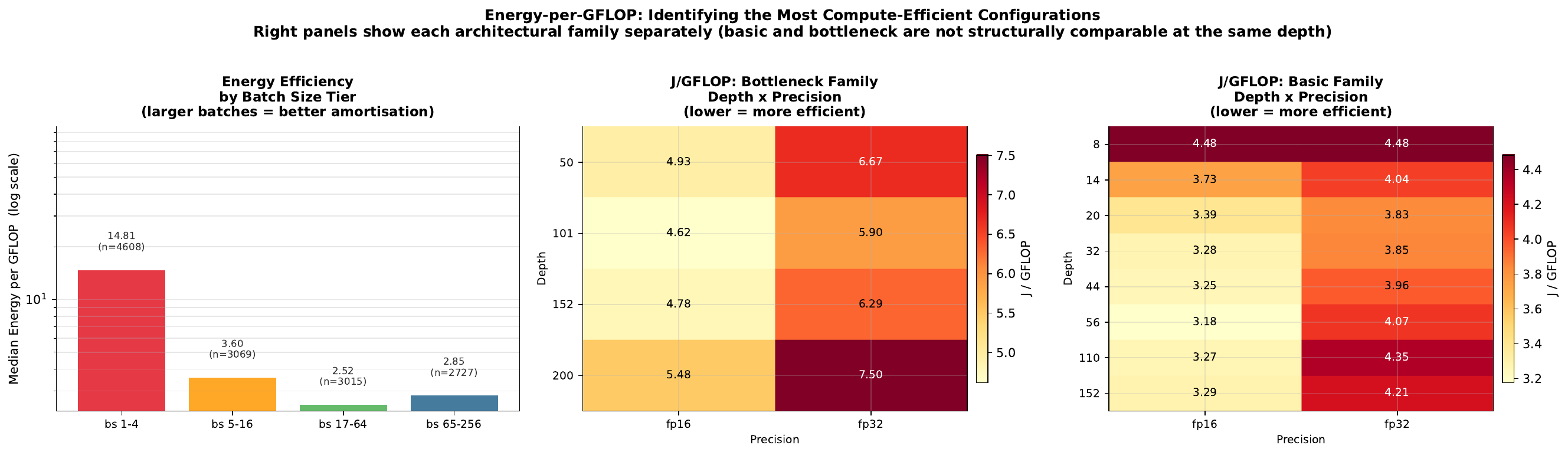}
    \caption{Energy per GFLOP (J/GFLOP) across three dimensions. Left: by batch-size tier (log scale). Right two panels: depth $\times$ precision heatmaps for the bottleneck family (center) and basic family (right) separately, since the two families cover different depth ranges and are not structurally comparable at the same depth value. Lower J/GFLOP indicates higher compute efficiency.}
    \label{fig:eff}
\end{figure*}

\subsection{Energy Characterization}
\label{sec:char}
 
Table~\ref{tab:knob_ranges} summarizes the median energy range across each design knob, with latency and memory shown for comparison. Figure~\ref{fig:char} visualizes the per-knob relationships, with the basic family (depths 8--152) and bottleneck family (depths 50--200) shown as separate series in every panel. 
 
\begin{table}[t]
\centering
\caption{Median range and max/min ratio per design knob, separated by architectural family.}
\label{tab:knob_ranges}
\setlength{\tabcolsep}{4pt}
\begin{tabular}{lrrrrrr}
\toprule
 & \multicolumn{2}{c}{\textbf{Energy (J)}} & \multicolumn{2}{c}{\textbf{Latency (ms)}} & \multicolumn{2}{c}{\textbf{Memory (MB)}} \\
\cmidrule(lr){2-3}\cmidrule(lr){4-5}\cmidrule(lr){6-7}
\textbf{Knob} & Range & $\times$ & Range & $\times$ & Range & $\times$ \\
\midrule
\multicolumn{7}{l}{\textit{Basic block (depths 8--152)}} \\
\midrule
Batch Size  & 27.8--440.6  & 15.9 & 4.81--24.32 & 5.1 & 29.8--438.3  & 14.7 \\
Depth       & 10.0--138.0  & 13.8 & 1.50--17.44 & 11.6 & 44.7--102.5 & 2.3 \\
Width Mult. & 25.5--185.6  & 7.3  & 4.31--15.64 & 3.6 & 19.7--291.8  & 14.8 \\
Input Size  & 36.8--102.6  & 2.8  & 5.56--10.12 & 1.8 & 41.0--125.3  & 3.1 \\
Early DS    & 51.6--72.6   & 1.4  & 6.55--7.85  & 1.2 & 64.1--82.8   & 1.3 \\
Precision   & 59.4--63.0   & 1.1  & 7.01--7.21  & 1.0 & 56.2--93.2   & 1.7 \\
\midrule
\multicolumn{7}{l}{\textit{Bottleneck block (depths 50--200)}} \\
\midrule
Batch Size  & 65.7--742.7  & 11.3 & 10.88--42.19 & 3.9 & 38.0--574.6 & 15.1 \\
Depth       & 44.7--150.6  & 3.4  & 5.58--20.05  & 3.6 & 96.4--122.7 & 1.3 \\
Width Mult. & 64.3--299.8  & 4.7  & 10.74--21.65 & 2.0 & 24.6--363.8 & 14.8 \\
Input Size  & 78.6--172.1  & 2.2  & 12.18--16.79 & 1.4 & 57.5--206.6 & 3.6 \\
Early DS    & 95.9--129.4  & 1.3  & 13.51--15.18 & 1.1 & 93.2--136.2 & 1.5 \\
Precision   & 105.4--111.7 & 1.1  & 13.48--15.11 & 1.1 & 87.4--147.4 & 1.7 \\
\bottomrule
\multicolumn{7}{l}{\footnotesize DS\,=\,Early Downsample; $\times$\,=\,max/min ratio of per-knob medians.}
\end{tabular}
\end{table}
 
Depth and batch size are the two dominant energy drivers within the basic family, spanning $13.8\times$ and $15.9\times$ respectively. The bottleneck family shows a narrower depth range ($3.4\times$) because all bottleneck depths are architecturally deep and compute-intensive; batch size remains dominant at $11.3\times$. Both families show that batch size widens the energy-to-latency gap sharply---basic: $15.9\times$ vs.\ $5.1\times$, bottleneck: $11.3\times$ vs.\ $3.9\times$---pointing to bandwidth overhead that grows faster than execution time.
 
Width multiplier shows a similar asymmetry: within the basic family, energy scales $7.3\times$ while memory scales $14.8\times$; within the bottleneck family, energy scales $4.7\times$ while memory scales $14.8\times$. In both cases memory grows twice as fast as energy, since wider networks allocate proportionally larger activation maps. Width multiplier is therefore the primary lever for memory-constrained deployment.
 
The separation of families confirms that the block-type energy difference (bottleneck family draws $1.8\times$ more energy than basic on average) is visible within every panel, indicating that the bottleneck family draws more energy across all levels of batch size, width, and input size. Within each family, FP32 vs.\ FP16 produces only a $1.1\times$ energy difference---the smallest of all knobs---while causing a $1.7\times$ memory difference. Switching to FP16 is therefore primarily a memory optimization, not an energy one.
 
Across all panels, the wide IQR bands shown in Figure~\ref{fig:char} indicate that the same knob setting can produce very different energy levels depending on other configuration choices, motivating a model that considers all design parameters jointly rather than treating each in isolation.

\subsection{Feature Correlation Analysis}
\label{sec:corr}

Table~\ref{tab:correlation} shows Pearson $|r|$ between each feature and the three targets, sorted by $|r|$ with energy (top 10 shown). FLOPs are widely used as a compute-cost proxy~\cite{yang2017designing}, and correlate strongly with energy ($|r|\,{=}\,0.91$), but are not sufficient alone. Precision, which does not change the number of operations, causes up to $1.37\times$ variation in energy efficiency (J/GFLOP) for the same architecture (Section~\ref{sec:eff}), confirming that FLOPs-based estimation misses factors that meaningfully affect energy in practice.

\begin{table}[h]
\centering
\caption{Pearson $|r|$ between each feature and the three prediction targets, sorted by $|r|$ with energy (top 10).}
\label{tab:correlation}
\setlength{\tabcolsep}{5pt}
\begin{tabular}{lrrr}
\toprule
\textbf{Feature} & \textbf{Energy} & \textbf{Latency} & \textbf{Memory} \\
\midrule
FLOPs                   & 0.91 & 0.89 & 0.60 \\
Total Activation (MB)   & 0.89 & 0.90 & 0.80 \\
Max Activation (MB)     & 0.68 & 0.69 & 0.97 \\
FLOPs / Param           & 0.44 & 0.44 & 0.54 \\
Batch Size              & 0.38 & 0.37 & 0.46 \\
SM Utilization (\%)     & 0.37 & 0.36 & 0.46 \\
GPU Temp ($^\circ$C)    & 0.37 & 0.36 & 0.35 \\
Memory Utilization (\%) & 0.35 & 0.34 & 0.57 \\
\# Parameters           & 0.32 & 0.33 & 0.21 \\
Arith.\ Intensity       & 0.32 & 0.30 & 0.24 \\
\bottomrule
\end{tabular}
\end{table}


Memory prediction is dominated by max activation size ($|r|\,{=}\,0.97$): peak memory is governed by the largest intermediate tensor in the forward pass rather than total operation count, consistent with the width-multiplier asymmetry in Section~\ref{sec:char}.

SM utilization shows a moderate correlation with energy ($|r|\,{=}\,0.37$), which reflects that computationally intensive configurations tend to both stress the GPU and consume more energy. This supports the characterization of hardware demand as a workload property.

\subsection{Energy--Latency Divergence}
\label{sec:crossfactor}

Figure~\ref{fig:crossfactor} bins configurations by batch-size tier and GPU SM utilization level, and reports mean energy and latency per bin. The rightmost panel shows the energy compounding ratio per batch tier. Among configurations in the high-SM-utilization, large-batch tier, energy scales $35.4\times$ relative to the low-SM, small-batch baseline, while latency scales only $11.2\times$ under the same conditions---a $3\times$ divergence. A practitioner who profiles only latency would underestimate the true energy overhead by a factor of three or more in the regimes most relevant to deployment.

The divergence is not uniform: the High-SM\,/\,Low-SM energy ratio grows monotonically from $3.29\times$ at small batches to $35.4\times$ at large batches, while latency shows no comparable amplification. This suggests that energy may capture overheads associated with memory bandwidth pressure and sustained hardware activity that latency alone does not reflect, and that the two metrics respond differently even to the same intervention---confirming that latency does not fully represent energy.



\subsection{Energy Efficiency Analysis}
\label{sec:eff}

Figure~\ref{fig:eff} reports energy per GFLOP (J/GFLOP) as a normalized efficiency metric. Precision causes up to $1.37\times$ variation in J/GFLOP for the same architecture: within the bottleneck family, depth\,200 achieves 5.48\,J/GFLOP in FP16 vs.\ 7.50 in FP32, while depth\,44 basic shows 3.25 vs.\ 3.96. Since FLOPs do not change with precision for the same architecture, this gap reflects differences in how efficiently the hardware executes different numeric formats---a factor invisible to any FLOPs-based model.

Batch size has an even stronger effect on compute efficiency. J/GFLOP falls from 14.81 at bs\,1--4 to 2.52 at bs\,17--64---a $5.9\times$ improvement---before plateauing at 2.85 for bs\,65--256, consistent with the GPU approaching full occupancy beyond bs\,17. Running at very small batch sizes therefore incurs a large per-GFLOP energy penalty that is independent of the model architecture.

The right two panels of Figure~\ref{fig:eff} confirm this non-linearity per family: within the bottleneck family, J/GFLOP reverses from 5.17 (depth\,152) to 7.50 (depth\,200) despite increasing FLOPs; the basic family shows a U-shaped pattern with a minimum around depth\,44--56. Neither pattern is predictable from FLOPs alone, motivating CARB, the data-driven ensemble model described in Section~\ref{sec:model}.

\subsection{Cross-GPU Generalizability}
\label{sec:crossgpu}

\begin{figure}[t]
\centering
  \includegraphics[width=0.9\linewidth]{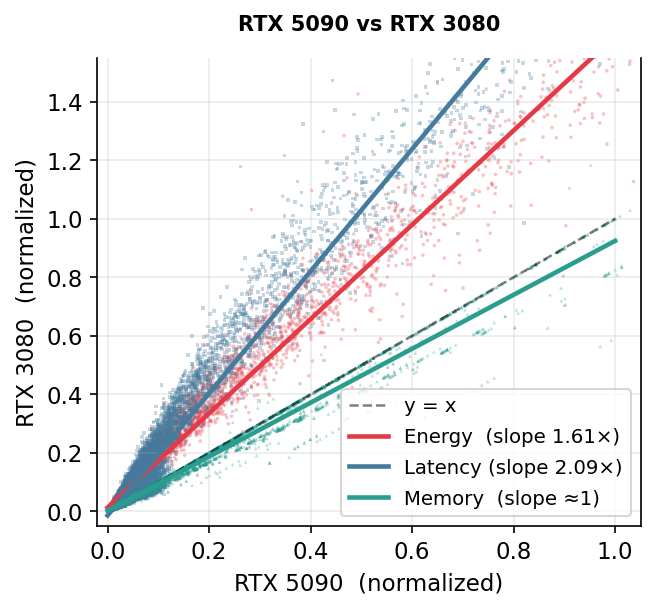}
  \caption{Per-configuration scatter: RTX\,5090 ($x$-axis) vs.\ RTX\,3080 ($y$-axis) for energy, latency, and memory. Points on the diagonal indicate identical cost. Linear fits characterize the cross-GPU scaling relationship for each target.}
  \label{fig:gpu_scatter}
\end{figure}

We profile the same configurations on an RTX\,3080 and compare against the primary RTX\,5090 measurements. Figure~\ref{fig:gpu_scatter} shows per-configuration scatter plots with linear fits for all three targets.

The strong linear structure visible across all three panels confirms that architectural ranking is preserved across the two tested GPU platforms: configurations that are energy-expensive on the RTX\,5090 remain expensive on the RTX\,3080, and vice versa. This rank invariance makes cross-platform design-space screening feasible---a model trained on one GPU can reliably identify which candidate architectures are worth profiling on another, without re-running the full benchmark.

Energy and latency, however, do not transfer with a fixed multiplier. The energy fit ($y\,{=}\,1.61x\,{+}\,52.8$, slope\,$>\,1$) shows that the RTX\,3080 is consistently more expensive, with the absolute gap widening as workload intensity grows. The latency fit ($y\,{=}\,2.09x\,{-}\,2.7$) follows a similar pattern. A practitioner who applies a single scale factor from one GPU to another will therefore systematically mis-estimate costs at either end of the workload range, making per-GPU models necessary for accurate prediction.

Memory behaves differently from both energy and latency. The memory fit ($y\,{=}\,0.92x\,{+}\,6.7$, slope\,$\approx\,1$) lies near the $y\,{=}\,x$ diagonal with a tight cluster, suggesting that peak memory is primarily determined by model architecture and batch size rather than GPU hardware characteristics across our two tested platforms. A single memory model trained on either GPU therefore transfers well to the other, without requiring separate retraining. These three findings directly motivate the two-stage framework in Section~\ref{sec:dse}.
\section{Prediction Framework (CARB)}
\label{sec:model}

\begin{figure*}[t]
\centering
\begin{tikzpicture}[
    font=\footnotesize,
    every node/.style={inner sep=3pt},
    mbox/.style={draw, rounded corners=3pt, minimum height=0.45cm,
                 minimum width=1.3cm, align=center},
    feat/.style={mbox, fill=blue!10, draw=blue!40},
    model/.style={mbox, fill=teal!15, draw=teal!50},
    stage/.style={mbox, fill=orange!12, draw=orange!50},
    res/.style={mbox, fill=green!12, draw=green!50, minimum width=1.4cm},
    arr/.style={->, >=stealth, thick},
    darr/.style={->, >=stealth, thick, dashed, gray!70},
    lbl/.style={font=\footnotesize\itshape, gray},
]

\node[feat]                              (arch) at (0,  0.25) {Arch. Features};
\node[feat, minimum width=1.3cm]         (run)  at (0, -0.4) {Runtime (optional)};
\node[draw=gray!55, dashed, rounded corners=4pt,
      fit=(arch)(run), inner sep=5pt]    (features) {};

\node[model] (mem) at (2.5, 0) {Memory\\Model};
\node[model, right=0.7cm of mem] (eng)   {Energy\\Model};
\node[model, right=0.7cm of eng] (lat)   {Latency\\Model};

\coordinate (ftop) at ([yshift=0.45cm]features.north);
\draw[arr, blue!60] (features.north) -- (ftop) -| (mem.north);
\draw[arr, blue!60] (features.north) -- (ftop) -| (eng.north);
\draw[arr, blue!60] (features.north) -- (ftop) -| (lat.north);

\draw[arr, teal!70] (mem) -- node[above, font=\tiny]{cascade} (eng);
\draw[arr, teal!70] (eng) -- node[above, font=\tiny]{cascade} (lat);


\draw[gray!35, dashed, thick] (7.45, 0.9) -- (7.45, -1.1);

\node[feat, right=0.5cm of lat]   (cand) {Candidate\\Grid};
\node[stage, right=0.9cm of cand] (s1)  {\textbf{Stage 1}\\{\scriptsize RTX 5090 rank}};
\node[stage, right=0.55cm of s1]   (s2)  {\textbf{Stage 2}\\{\scriptsize RTX 3080 budget}};
\node[res,   right=0.55cm of s2]   (par) {Pareto\\Shortlist};

\draw[arr] (cand) -- node[above, font=\tiny]{all configs} (s1);
\draw[arr] (s1)   -- node[above, font=\tiny]{top-$K$} (s2);
\draw[arr] (s2)   -- node[above, font=\tiny]{Pareto} (par);

\draw[arr, gray!55]
    (mem.south) |- ([yshift=-0.55cm]mem.south -| s1.south) -- (s1.south);
\draw[arr, gray!55]
    (eng.south) |- ([yshift=-0.55cm]eng.south -| s1.south) -- (s1.south);
\draw[arr, gray!55]
    (lat.south) |- ([yshift=-0.55cm]lat.south -| s1.south)
    node[below, font=\small, gray, pos=0.5,xshift=-2cm]{CARB predictions} -- (s1.south);

\node[lbl, above=0.15cm of s1, xshift=0.9cm]
    {\textbf{Two-Stage Deployment Screening}};

\end{tikzpicture}
\caption{CARB framework overview. \textit{Left}: both architectural and optional runtime features feed into all three cascade models (memory $\to$ energy $\to$ latency). \textit{Right}: CARB predictions drive two-stage screening---Stage\,1 (RTX\,5090 ranking) shortlists candidates; Stage\,2 (RTX\,3080 budget filter) extracts the Pareto-optimal configurations.}
\label{fig:framework}
\end{figure*}
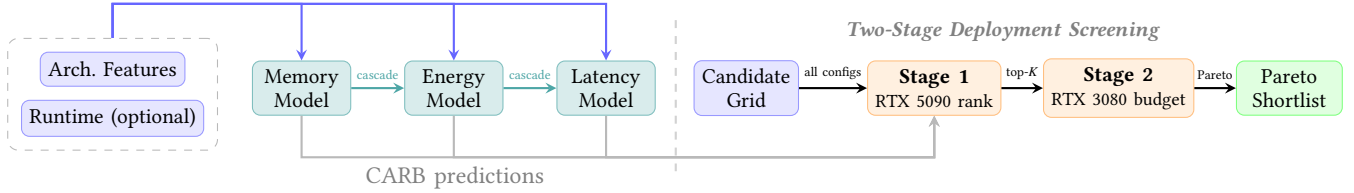

Building on the characterization findings in Section~\ref{sec:analysis}, we develop CARB (Context-Aware Regime-corrected Blended ensemble), a multi-target framework that jointly estimates peak memory, energy, and latency for a given ResNet configuration, built around three principles: physical coupling between targets, hardware-load interaction features, and regime-specific residual correction. Figure~\ref{fig:framework} provides an overview
of the complete framework: the left panel shows the cascade
prediction pipeline, and the right panel shows how CARB
predictions drive the two-stage deployment screening
described in Section~\ref{sec:dse}.

\subsection{Feature Engineering}
To capture compounding hardware-load effects that additive features cannot represent, we construct multiplicative interaction features from architectural parameters and live GPU telemetry:
\begin{itemize}
    \item $\text{batch\_x\_sm} = \text{batch\_size} \times \text{avg\_sm\_util}$
    \item $\text{flops\_x\_sm} = \log(1+\text{flops}) \times \text{avg\_sm\_util}$
    \item $\text{act\_x\_mem} = \log(1 + \text{total\_act\_MB}) \times \text{avg\_mem\_util}$
    \item $\text{util\_product} = \text{avg\_sm\_util} \times \text{avg\_mem\_util}$
\end{itemize}
Heavy-tailed raw columns (FLOPs, parameter count, activation size) are log-transformed to compress scale and improve tree split quality. Categorical variables (block type, precision) are one-hot encoded. In total, the feature matrix combines static architectural descriptors with dynamic hardware context, giving the model a direct handle on the joint effect of model design and hardware state. These telemetry-based interaction features are used in CARB's full-telemetry mode; Section~\ref{sec:ablation} demonstrates that an architecture-only configuration achieves comparable accuracy when live instrumentation is unavailable.

\subsection{Stratified Data Split}

The dataset is partitioned 70/15/15 into train, validation, and test sets using stratified sampling. The stratification key is a joint combination of batch size tier (small $\leq 4$, medium $\leq 16$, large $\leq 128$), hardware stress flag (SM or memory utilization $> 65\%$), and depth tier (shallow, mid, deep). This ensures that edge-case configurations like high-load and small-batch 
regimes are proportionally represented in every partition. To verify generalization beyond interpolation, a leave-one-batch-tier-out evaluation --- training on batch size $> 8$ and testing on batch size $\leq 8$ entirely withheld from training --- yields $R^2 = 0.956$, $0.991$, and $0.968$ (MAE $=$ 6.73\,MB, 7.78\,J, 1.11\,ms) for memory, energy, and latency 
respectively, confirming that results are not an artifact of configuration leakage across the train/test boundary.

\subsection{Specialist Ensemble with Cascade Prediction}

The core of CARB is a blended ensemble of three structurally diverse base learners trained per target:

\begin{itemize}
\item \textbf{XGBoost} ~\cite{chen2016xgboost} (800 trees, depth 9) — level-wise boosting, strong on dense tabular interactions
\item \textbf{LightGBM} ~\cite{ke2017lightgbm} (800 trees, 127 leaves) — histogram-based leaf-wise growth, captures fine-grained feature splits efficiently
\item \textbf{ExtraTrees} ~\cite{geurts2006extratrees} (400 trees) — fully random split thresholds, maximally decorrelated from the boosting models
\end{itemize}

Combining boosting with bagging produces a specialist in which prediction errors are less correlated than those of individual models. The three predictions are blended with validation-calibrated weights 
found by grid search on the validation set (Table~\ref{tab:blend_weights}). 
Energy is dominated by LightGBM (weight 0.5), whose leaf-wise tree growth effectively captures nonlinear interactions in the feature space. Latency is dominated by ExtraTrees (weight 0.5) --- 
that random splits perform as well as optimized ones for latency but 
not energy suggests the latency surface is more regularly structured. 
Peak memory weights are approximately balanced, consistent with a 
largely linear model-size surface.
Critically, the three targets are not predicted independently. The empirical dependency structure — where memory, energy, and latency exhibit correlated behavior — is captured via cascade prediction:
\vspace{-0.03in}
\begin{align}
    \hat{y}_{\mathrm{mem}} &= f_{\mathrm{mem}}(\mathbf{x}) \\
    \hat{y}_{\mathrm{energy}} &= f_{\mathrm{energy}}(\mathbf{x},\, \hat{y}_{\mathrm{mem}}) \\
    \hat{y}_{\mathrm{latency}} &= f_{\mathrm{latency}}(\mathbf{x},\, \hat{y}_{\mathrm{mem}},\, \hat{y}_{\mathrm{energy}})
\end{align}

Each downstream model receives upstream predictions as soft priors, exploiting inter-target dependencies that independent regression cannot capture.
\begin{table}[h]
  \centering
  \caption{Validation-calibrated specialist blend weights per target.
    Weights are (XGBoost, LightGBM, ExtraTrees).}
  \label{tab:blend_weights}
  \small
  \begin{tabular}{lccc}
    \toprule
    \textbf{Target} & \textbf{XGBoost} & \textbf{LightGBM} & \textbf{ExtraTrees} \\
    \midrule
    \text{peak\_memory\_MB} & 0.3 & 0.4 & 0.3 \\
    \text{energy\_J}        & 0.3 & 0.5 & 0.2 \\
    \text{latency\_ms}      & 0.2 & 0.3 & 0.5 \\
    \bottomrule
  \end{tabular}
\end{table}

\subsection{Residual Correctability Analysis}
Before training any corrector, we examine whether the specialist's residuals contain learnable structure. For each target, we compute the Spearman correlation between residuals and hardware utilization features on the training set, restricted to the high-stress regime (avg\_SM\_util $> 65\%$). The results are:
\begin{itemize}
    \item \text{peak\_memory\_MB}: $\rho = 0.53$ with avg SM utilization
    \item \text{energy\_J}: $\rho = 0.004$ with avg SM utilization
    \item \text{latency\_ms}: $\rho = 0.017$ with avg SM utilization
\end{itemize}
For energy and latency, near-zero correlation indicates a hardware-stress corrector would learn noise. Residual standard deviation analysis identifies the low-batch regime (batch size $\leq 8$) as the common axis of elevated, structured error across all three targets.

\subsection{Regime-Specific Residual Correctors}
\label{sec:correctors}
Based on the correctability analysis, a LightGBM corrector is trained per target on the specialist's residuals, restricted to the low-batch regime. Training on only this regime prevents the corrector from fitting noise in samples where no correctable structure exists. The corrected prediction is:
\vspace{-0.1in}
\begin{equation}
\hat{y}_{\text{corrected}} = \hat{y}_{\text{specialist}} + \lambda \cdot \hat{r}
\end{equation}
where $\hat{r}$ is the corrector's predicted residual and $\lambda = 0.9$ is a damping factor that prevents overcorrection.
The final CARB prediction pipeline is therefore:
\begin{equation}
\hat{y}_{\text{CARB}} = \hat{y}_{\text{specialist}} + 0.9 \cdot \hat{r}_{\text{corrector}}
\end{equation}
All targets and all predictions are produced in log space (via log1p transformation) and inverse-transformed via expm1 for evaluation and reporting.

To contextualize CARB's accuracy, Table~\ref{tab:baselines} compares
against two baselines. A FLOPs-only linear regression achieves $R^2 < 0.38$ across all
targets, confirming that compute volume alone is insufficient for
reliable inference cost prediction. A latency-as-energy proxy
achieves high $R^2$ but a mean absolute error of 71.37\,J ---
demonstrating that latency correlates with energy but cannot
substitute for it in absolute terms, making it unsuitable for
budget-constrained deployment decisions. CARB reduces energy MAE
to 25.98\,J, a 63.6\% reduction over the latency proxy.

\begin{table}[h]
\centering
\caption{Baseline comparison on the overall test set. B1: FLOPs-only
linear regression. B2: latency-as-energy proxy (linear fit on
measured latency; memory and latency not applicable).
CARB: full pipeline.}
\label{tab:baselines}
\setlength{\tabcolsep}{4pt}
\resizebox{\columnwidth}{!}{%
\begin{tabular}{llrrr}
\toprule
\textbf{Model} & \textbf{Metric}
    & \textbf{Peak Memory} & \textbf{Energy} & \textbf{Latency} \\
\midrule
\multirow{2}{*}{B1: FLOPs-only}
    & $R^2$  & 0.379 & 0.279 & 0.233 \\
    & MAE    & 161.11\,MB & 324.07\,J & 18.32\,ms \\
\midrule
\multirow{2}{*}{B2: Latency proxy}
    & $R^2$  & --- & 0.993 & --- \\
    & MAE    & --- & 71.37\,J & --- \\
\midrule
\multirow{2}{*}{\textbf{CARB (full)}}
    & $R^2$  & \textbf{0.997} & \textbf{0.993} & \textbf{0.992} \\
    & MAE    & \textbf{6.93}\,MB & \textbf{25.98}\,J & \textbf{1.58}\,ms \\
\bottomrule
\end{tabular}}
\end{table}
\section{Explainability and Insights}

Beyond predictive accuracy, CARB's internal structure yields interpretable findings that both validate its design decisions and confirm the characterization findings of Section~\ref{sec:analysis}.

\subsection{Feature Importance}
\label{sec:feature_importance}

\begin{figure*}[t]
  \centering
  \includegraphics[width=\textwidth]{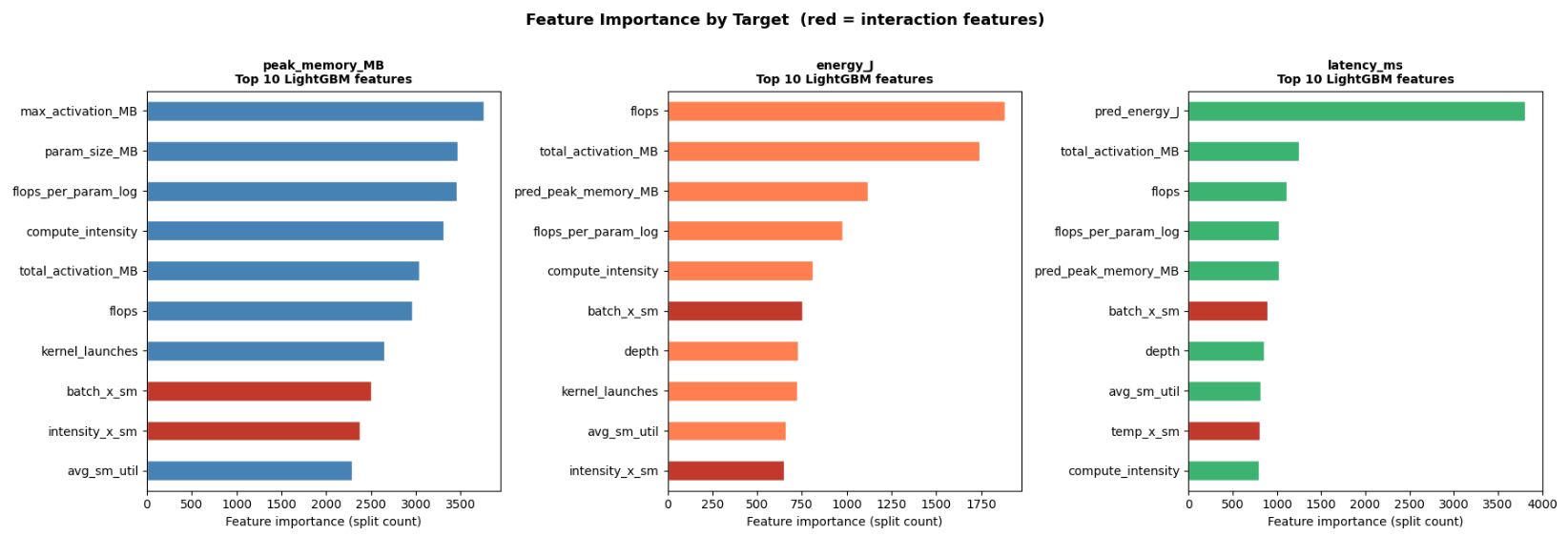}
  \caption{Top-10 LightGBM feature importances (split count) per
           prediction target. Interaction features
           ($f_{\text{arch}} \times f_{\text{hw}}$) are highlighted
           in red. Peak memory is dominated by raw architectural size
           features; interaction features only enter from rank~8.
           Energy and latency additionally show strong contributions
           from cascade predictions (\text{pred\_peak\_memory\_MB},
           \text{pred\_energy\_J}), with \text{pred\_energy\_J}
           ranking first for latency.}
  \label{fig:feature_importance}
\end{figure*}

Figure~\ref{fig:feature_importance} shows the top-10 LightGBM
split-based importances per specialist model.  The rankings reveal qualitatively different drivers per target, with interaction features, hardware telemetry, and cascade predictions varying substantially in role.

\paragraph{Peak memory}
The top six features for \text{peak\_memory\_MB} are consistently dominated by raw architectural or derived model-size quantities that directly reflect tensor and parameter scale, including \text{max\_activation\_MB}, \text{param\_size\_MB}, \text{flops\_per\_param\_log}, \text{compute\_intensity}, \text{total\_activation\_MB}, and \text{flops}. Interaction features and hardware telemetry signals such as \text{batch\_x\_sm} and \text{avg\_sm\_util} only enter from rank~8 onwards.  This reflects an empirical regularity: peak memory allocation is determined by activation tensors and parameter buffers fixed by the model graph before any kernel executes.

\paragraph{Inference energy}
Energy is led by raw compute and activation volume (\text{flops},
\text{total\_activation\_MB}) followed by \text{flops\_per\_param\_log}
and \text{compute\_intensity}.  Interaction features
(\text{batch\_x\_sm}, \text{intensity\_x\_sm})
appear at ranks 6 and 10 respectively, providing additional signal beyond the dominant architectural features. \text{avg\_sm\_util} appears within the top-10 for all three targets, but enters later for memory (rank 10) than for energy (rank 9) or latency (rank 8), confirming that hardware state plays a secondary role relative to architectural features across all targets.


\paragraph{Inference latency}
For latency, \text{pred\_energy\_J} --- the upstream cascade prediction --- ranks first, ahead of every raw architectural feature. \text{total\_activation\_MB}, \text{flops}, and \text{flops\_per\_param\_log} follow at ranks 2--4, with \text{pred\_peak\_memory\_MB} appearing at rank~5. Interaction features (\text{batch\_x\_sm}, \text{temp\_x\_sm}, \text{compute\_intensity}) are consistently present but secondary. The role of cascade predictions for latency is discussed further in Section~\ref{sec:cascade_insight}.

\paragraph{Cross-target feature overlap}
Three features appear in the top six for all three targets:
\text{flops}, \text{total\_activation\_MB}, and
\text{flops\_per\_param\_log}. These capture total compute volume,
total memory footprint, and parameter efficiency --- the three
dimensions that jointly determine the cost of a forward pass
regardless of which specific cost metric is being predicted.
Table~\ref{tab:feature_overlap} summarizes the cross-target top-5 membership.

These rankings independently reproduce the principal findings of Section~\ref{sec:analysis} without any direct exposure to that analysis during training. The consistent presence of \text{batch\_x\_sm}, \text{intensity\_x\_sm}, and \text{flops\_x\_sm} in the energy and latency top-10 reflects the super-linear energy compounding under SM contention observed in Section~\ref{sec:crossfactor}, while the near-exclusive dominance of architectural features for memory mirrors the cross-platform slope ${\approx}1$ finding of Section~\ref{sec:crossgpu}. In both cases, the model's learned weights converge on the same physical structure the empirical characterization identifies directly.

\begin{table}[t]
  \centering
  \caption{Top-5 feature membership across prediction targets.
    $\checkmark$ indicates the feature ranks within the top 5 for
    that target's LightGBM specialist.
    $\dagger$ marks interaction features; $\ddagger$ marks cascade
    (upstream prediction) features. Memory indicates peak memory.}
  \label{tab:feature_overlap}
  \small
  \begin{tabular}{lccc}
    \toprule
    \textbf{Feature} & \textbf{Memory} & \textbf{Energy} & \textbf{Latency} \\
    \midrule
    \text{flops}                        & \checkmark & \checkmark & \checkmark \\
    \text{total\_activation\_MB}        & \checkmark & \checkmark & \checkmark \\
    \text{flops\_per\_param\_log}       & \checkmark & \checkmark & \checkmark \\
    \text{max\_activation\_MB}          & \checkmark &            &            \\
    \text{param\_size\_MB}              & \checkmark &            &            \\
    \text{compute\_intensity}           & \checkmark & \checkmark &            \\
    \text{pred\_peak\_memory\_MB}$^\ddagger$ &       & \checkmark & \checkmark \\
    \text{pred\_energy\_J}$^\ddagger$   &            &            & \checkmark \\
    \bottomrule
  \end{tabular}
\end{table}

\subsection{Memory is Architecturally Determined}
\label{sec:memory_insight}

The near-exclusive dominance of raw architectural features for peak memory prediction has a direct practical implication: memory consumption can be reliably estimated from the model graph alone, without any live hardware observation.  This is consistent with the cross-GPU empirical finding in Section~\ref{sec:crossgpu}, where peak memory follows a cross-GPU linear fit with slope $\approx 1$, confirming that memory allocation is preserved across the two tested GPU platforms.

\subsection{Runtime Feature Ablation}
\label{sec:ablation}

The feature importance analysis in Section~\ref{sec:memory_insight}
suggests that hardware telemetry contributes little additional
information for memory prediction once architectural size features
are known. We test this hypothesis through a controlled
feature ablation study, isolating the contribution of
runtime feature groups across all three targets.

Three model configurations are evaluated on identical
train/validation/test splits. \textbf{M1 (Full)} retains the
complete feature set including live GPU telemetry.
\textbf{M2 (No Util)} removes utilization features
(\text{avg\_sm\_util}, \text{avg\_mem\_util}, and derived
interaction terms) to assess the predictive value of dynamic utilization metrics. \textbf{M3 (Arch-only)} additionally removes
\text{gpu\_temp\_C} and \text{kernel\_launches}, leaving primarily static architectural and workload descriptors. All other pipeline components,
including ensemble structure, cascade ordering, and blend
weights, are held fixed. Table~\ref{tab:ablation} reports
test-set $R^2$ values across all evaluation regimes.

Removing all telemetry features reduces $R^2$ by at most 0.0012 in the
overall regime across all three targets. M3 achieves $R^2$ of 0.9963,
0.9892, and 0.9904 for memory, energy, and latency respectively ---
within noise of the full-telemetry model --- confirming that hardware
telemetry is not required for reliable prediction of any of the three
cost targets.

The per-regime breakdown reveals one notable anomaly. Low-batch
peak memory and high-load energy prediction achieve marginally
higher $R^2$ values for M2 than for M1. This suggests that
utilization features may introduce mild noise in these regimes,
where SM utilization is more variable and less predictive than
in medium-to-large batch settings. Although small, the effect
is consistent across runs.

The practical implication is significant. CARB therefore supports
two operational modes: a \textbf{full-telemetry mode} that
incorporates live GPU utilization and thermal features for maximum
accuracy in instrumented environments, and a
\textbf{telemetry-free mode} that relies on static architectural
parameters alone when live GPU instrumentation is unavailable.
The ablation results confirm that the telemetry-free mode achieves
$R^2$ within 0.0012 of the full-telemetry model across all three
targets, making it a viable configuration for design-stage
screening before hardware access is available.

\begin{table}[t]
\centering

\caption{
Test-set $R^2$ values under three feature configurations across
evaluation regimes. $\Delta$ denotes the difference in $R^2$ between the full-telemetry model (M1) and the architecture-only model (M3).
}

\label{tab:ablation}

\resizebox{\columnwidth}{!}{%

\begin{tabular}{llcccc}
\hline

\textbf{Target} &
\textbf{Regime} &
\textbf{M1 (Full)} &
\textbf{M2 (No Util)} &
\textbf{M3 (Arch-only)} &
\textbf{$\Delta$ (M1$\rightarrow$M3)} \\

\hline

Peak Memory
& Overall
& 0.9967
& 0.9964
& 0.9963
& -0.0004 \\

& High-load
& 0.9947
& 0.9939
& 0.9939
& -0.0008 \\

& Low-batch
& 0.9986
& \textbf{0.9990}
& 0.9989
& +0.0003 \\

\hline

Energy
& Overall
& 0.9900
& 0.9897
& 0.9892
& -0.0008 \\

& High-load
& 0.9893
& \textbf{0.9894}
& 0.9886
& -0.0007 \\

& Low-batch
& 0.9941
& \textbf{0.9942}
& 0.9936
& -0.0005 \\

\hline

Latency
& Overall
& 0.9916
& 0.9909
& 0.9904
& -0.0012 \\

& High-load
& 0.9910
& 0.9904
& 0.9900
& -0.0010 \\

& Low-batch
& 0.9717
& 0.9683
& 0.9672
& -0.0045 \\

\hline
\end{tabular}
}

\begin{tablenotes}
\small
\item Bold entries indicate regimes where removing utilization features 
marginally \emph{improves} $R^2$ over the full model.
\end{tablenotes}

\end{table}

\subsection{Cascade as Physically Grounded Coupling}
\label{sec:cascade_insight}

The cascade design encodes the empirical dependency structure --- memory constrains energy, energy bounds latency --- by passing upstream predictions as input features to downstream models ~\cite{spyromitros2016multitarget}.  The feature
importance rankings provide the empirical
validation of this design: \text{pred\_peak\_memory\_MB} ranks
third for energy prediction, and \text{pred\_energy\_J} ranks
first for latency prediction, above all raw architectural
features in the dataset.

The latency result in particular is non-trivial.  It shows that once
the model has an estimate of how much energy a configuration will
consume, that single signal is more predictive of latency than
\text{flops}, \text{total\_activation\_MB}, or any interaction
term.  Energy integrates compute volume, memory traffic, SM utilization, and precision simultaneously --- suggesting it acts as a compressed summary of execution time determinants. The cascade propagates this signal directly, reducing the burden on the downstream model to reconstruct it from lower-level features.

\subsection{Low-Batch Residual Structure}
\label{sec:residual_structure}

A non-obvious finding from CARB's residual analysis is that the
hardest prediction regime across all three targets is not the
high-hardware-load regime (saturated SM or memory utilization) but the
low-batch regime ($\text{bs} \leq 8$).


Residual correlation reveals a sharp asymmetry: peak-memory
residuals correlate with SM utilization at $\rho = 0.53$ in the
high-load regime, indicating learnable structure that a hardware-stress
corrector can reduce. Energy and latency residuals, however, correlate
at $\rho = 0.004$ and $\rho = 0.017$ respectively --- effectively
zero --- so applying a hardware-stress corrector would learn noise
rather than signal. Instead, residual standard deviation analysis
identifies the low-batch regime as the shared axis of elevated,
structured prediction error across all three targets, motivating
the per-target low-batch correctors described in
Section~\ref{sec:correctors}.

This result is counterintuitive because low-batch configurations are computationally the simplest: small activations, minimal memory pressure, low GPU occupancy.  The explanation is that the specialist ensemble optimizes across the full batch distribution, implicitly weighting the majority medium-to-large-batch regime.  At very small batch sizes, kernel launch overhead, thermal transients, and driver scheduling latency constitute a larger fraction of total cost --- sources of variance that neither architectural features nor utilization telemetry can fully represent.

\section{Design-Space Exploration Framework}
\label{sec:dse}

During architecture design, identifying energy-efficient configurations
requires evaluating a large space of candidates across depth, width,
precision, and batch size---a process that, if done through hardware
profiling alone, scales poorly and demands hours of GPU time before
any deployment decision can be made.
CARB addresses this by serving as a pre-screening step: the full
candidate grid is scored through the CARB pipeline in seconds,
configurations predicted to exceed the deployment budget are eliminated,
and hardware profiling is reserved for only the Pareto-optimal
candidates that are worth measuring.

Both GPU-specific models achieve $R^2\,{\approx}\,0.99$: RTX\,5090 scores 0.993\,/\,0.992\,/\,0.997 and RTX\,3080 scores 0.995\,/\,0.995\,/\,0.998 for energy, latency, and memory respectively, establishing the prediction quality on which the screening decisions rest.

As illustrated in Figure~\ref{fig:framework} (right), \textbf{Stage~1 (Ranking-based screening)} uses a single GPU's CARB model to rank all candidates and narrow the pool, exploiting the rank-preservation property to eliminate configurations that are unlikely to be feasible on any target hardware without running any target-device inference.
\textbf{Stage~2 (GPU-specific prediction)} re-scores only the
shortlisted candidates with the target GPU's model, applying
device-specific energy and latency constraints to produce the final
Pareto frontier.
Memory predictions are shared across both stages, as memory transfers well across the two tested platforms (cross-GPU slope ${\approx}1$,
Section~\ref{sec:crossgpu}).

\subsection{Screening Workflow}

The screening process proceeds in four stages, forming the
implementation of Stage~2 once a shortlist has been produced by
the ranking-based Stage~1.

\begin{table*}[h]
\centering
\caption{Per-configuration screening validation results. CARB is evaluated on real held-out test feature vectors for each unique Pareto-feasible candidate. The budget threshold is 75\,J; $\checkmark$ indicates a correct budget classification.}
\label{tab:screening_validation}
\setlength{\tabcolsep}{4pt}
\begin{tabular}{cccccrrrrrc}
\toprule
\textbf{Block} & \textbf{Depth} & \textbf{Width} & \textbf{Prec.} & \textbf{BS}
    & \textbf{Pred.\ (J)} & \textbf{Actual (J)}
    & \textbf{Abs.\ Err.\ (J)} & \textbf{Pred.\ Mem.\ (MB)} & \textbf{Pred.\ Lat.\ (ms)}
    & \textbf{Decision} \\
\midrule
basic & 14 & 1.0 & fp32 & 1 & 11.72 & 15.69 & 3.96 & 53.02 & 7.91 & $\checkmark$ \\
basic & 14 & 2.0 & fp32 & 1 & 18.05 & 16.42 & 1.63 & 53.61 & 7.93 & $\checkmark$ \\
basic & 14 & 0.5 & fp32 & 1 & 12.86 & 14.73 & 1.87 & 52.98 & 7.91 & $\checkmark$ \\
\midrule
\multicolumn{7}{l}{\textit{Mean}} & 2.49 & & & 3/3 \\
\bottomrule
\end{tabular}
\end{table*}
\begin{enumerate}

\item \textbf{Enumerate.}
A grid of candidate configurations is constructed from the
architectural design space of interest, covering depth, width, block
type, precision, and batch size.
Runtime telemetry features (SM utilization, memory utilization, GPU
temperature) are fixed to training-set medians stratified by
batch-size tier, reflecting realistic hardware conditions without
requiring any GPU runs.

\item \textbf{Score.}
All candidates are passed through the full CARB pipeline, producing
predicted energy, peak memory, and latency for every configuration
in under one second.
This is the core practical value of CARB: exhaustive design-space
scoring at negligible cost compared to brute-force hardware profiling.

\item \textbf{Screen.}
Candidates are filtered against the deployment budget and ranked by
energy efficiency.
The Pareto-optimal frontier~\cite{deb2002nsga} is extracted, identifying configurations
that are not dominated on both energy and memory.
This directly surfaces the configurations worth deploying.

\item \textbf{Prioritize.}
Borderline configurations---those falling within one test-set MAE
of the budget boundary---are flagged for hardware profiling rather than accepted or rejected outright, guiding practitioners toward measuring exactly the cases where prediction uncertainty matters most.

\end{enumerate}

\subsection{Example: RTX 3080 Deployment Scenario}

We demonstrate the two-stage workflow targeting the NVIDIA RTX\,3080.
A deployment budget of 75\,J energy and 500\,MB peak memory is set. The candidate grid spans 4 depth values, 4 width multipliers, 4 block types, 3 precision formats, and 4 batch sizes, yielding 3,072 total configurations.
\textbf{Stage~1 (RTX\,5090 model, Ranking-based).}
All \num{3072} candidates are scored by the RTX\,5090 CARB model and
ranked by predicted energy.
The top-100 configurations (shortlist energy range 38.2--42.6\,J
on the RTX\,5090) are retained.
CARB predictions on the full \num{3072}-point candidate grid yield
a Spearman rank correlation of $\rho\,{=}\,0.95$ between the two
GPU models' energy rankings, confirming that configurations eliminated
at Stage~1 are highly unlikely to be feasible on the RTX\,3080.

\textbf{Stage~2 (RTX\,3080 model, GPU-specific).}
The 100 shortlisted configurations are re-scored by the RTX\,3080
model, accounting for GPU-specific energy and latency characteristics.
100\% pass the 75\,J / 500\,MB budget.
After Pareto extraction, \num{3072} candidates are reduced to
seven priority configurations---a 99.8\% reduction in profiling load. These span three unique architecture signatures across input resolution and early downsampling variants.

The three feasible configurations are shown in
Table~\ref{tab:screening_validation}.
All share basic-block depth 14, batch size 1,
and fp32 precision, differing in width multiplier. 
All achieve latency between 7.8--7.9\,ms with peak memory ranging from 52.98 to 53.61\,MB, representing energy-memory trade-off within the feasible region. 
A practitioner would select among these based on memory and throughput requirements.

Using the RTX\,5090 energy values directly to assess the 75\,J
RTX\,3080 budget (38--43\,J at Stage~1) would incorrectly classify
all shortlisted configurations as comfortably within budget, potentially
accepting configurations that are borderline or infeasible on the
actual deployment device.
Stage~2 is therefore essential for final budget decisions, even when
Stage~1 reliably narrows the candidate pool.

\subsection{Screening Validation}

To confirm that CARB's screening decisions are reliable, each
Pareto-feasible configuration is matched against real held-out test measurements from the RTX\,3080.
CARB is re-run on actual feature vectors---not the synthetic
architectural approximations used during screening---and predicted
energy is compared against measured values.

All matched configurations are correctly classified against the
deployment budget (3/3 correct budget decisions), with a mean absolute error of 2.49\,J and an RMSE of 2.70\,J.
The largest absolute error is 3.96\,J against a
75\,J ceiling, a difference that would not affect any deployment
decision. The results confirm that CARB reliably supports
design-stage screening where budget classification accuracy is
the relevant metric.

To assess classification reliability across the broader test set,
we evaluate CARB's budget decisions on all 1,828 RTX\,3080 test configurations against the 75\,J threshold. CARB achieves 95.8\% budget classification accuracy, with 2.1\% false accepts and 2.1\% false rejects. The symmetric error
distribution indicates no systematic bias toward over- or under-prediction at the budget boundary. False accepts --- configurations predicted feasible but actually exceeding the budget --- represent
the operationally critical error type; at 2.1\%, this rate is sufficiently low to support reliable pre-screening before
hardware profiling is committed to.
\section{Conclusion}

We presented a characterization-guided framework for CNN inference-cost prediction and deployment screening across 13\,419 configurations and two GPU platforms. The study demonstrates that common deployment proxies such as FLOPs or latency are insufficient for reliable energy-aware optimization, particularly under high computational demand where energy and latency decouple substantially. Results further indicate that transferability depends strongly on the prediction target: memory behavior generalizes across the two tested platforms more readily than energy or latency, motivating platform-specific modeling for accurate deployment estimation. 

Building on these findings, CARB achieves $R^2 \approx 0.99$ across all three targets while enabling rapid Pareto-guided pre-deployment filtering through a two-stage screening workflow. Together, the results demonstrate the importance of direct energy modeling for practical resource-aware CNN deployment.

\bibliographystyle{plain}
\bibliography{reference}

\end{document}